\documentclass[twocolumn,showpacs,amsmath,amssymb]{revtex4}
\usepackage{graphicx}
\usepackage{dcolumn}
\usepackage{bm}
\usepackage[dvips]{color}

\def\braket#1{\mid\!#1\rangle}

\begin{document}

\title{Charge Ordered RVB States in the Doped Cuprates}

\author{Huai-Xiang Huang$^1$, You-Quan Li$^1$, and Fu-Chun Zhang$^{1,2,3}$}

\affiliation{$^1$Zhejiang Institute of Modern Physics, Zhejiang University,
Hangzhou 310027, China\\
$^2$Department of Physics, Hong
Kong University, Hong Kong, China\\
$^3$Department of Physics, University of Cincinnati, Cincinnati, Ohio 45221}

\date{\today}%

\begin{abstract}
We study charge ordered d-wave resonating valence bond states
(dRVB) in the doped cuprates, and estimate the energies of these states
in a generalized
$t-J $ model by using a renormalized mean field theory.  The long
range Coulomb potential tends to modulate the charge density in
favor of the charge ordered RVB state. The possible relevance to the recently observed $4 \times
4$ checkerboard patterns in tunnelling conductance in high $T_c$ cuprates is discussed.
\end{abstract}


\maketitle

A number of recent STM experiments have shown spatial modulations
in tunnelling conductance in high $T_c$
cuprates~\cite{hoffman1,howald,hoffman2, mcelroy1,vershinin}. More
recent low temperature STM experiments have reported bias
independent modulations of period approximately $4 - 4.5a$ ($a$:
lattice constant) in the tunnelling conductance over a wide range
of energy on underdoped $Bi2212$~\cite{mcelroy2} and
$NaCCOC$~\cite{hanaguri}. Several theoretical proposals have been
put forward to interpret the observed checkerboard charge
ordering~\cite{franz,wang,chen,fu,anderson}. Chen {\it et
al.}~\cite{chen} have proposed that the modulations are due to
Cooper pair density wave. Fu {\it et al}~\cite{fu} have
examined the possibility of a soliton crystal in a generalized
Hubbard model including a nearest neighbor Coulomb repulsion and
an antiferromagnetic spin exchange coupling.
Anderson~\cite{anderson} has proposed an explicit wavefunction
describing a Wigner solid of holes embedded in a sea of $d$-RVB
state, and pointed out that the long-range Coulomb interaction
furnishes the energy gain and the stiffness of the hole
wavefunction opposes the deformation. The detailed calculations,
however,  have not been carried out in Ref.~\cite{anderson}.

In the present paper, we study the charge ordered
dRVB in the doped cuprates. We use a Gutzwiller
projected wavefunction with both BCS pairing and charge
ordering to describe the charge ordered state in
cuprates. Our approach is similar to the idea
outlined by Anderson~\cite{anderson}, who formulated the charge
ordering in a dRVB by a site-dependent fugacity,
which was introduced by Laughlin in the context of
Gutzwiller projected state in study of the Gossamer
superconductivity~\cite{laughlin,zhang03}.
Here we shall use the renormalized mean field
theory (RMFT) developed early~\cite{zhang88,vanilla}
to formulate charge ordering by site-dependent
renormalization factors and estimate the energies of
these states in the $t-J $ model. We show that the long range
Coulomb potential tends to modulate the charge density
in favor of the charge ordered RVB state and that the favorable
patterns for the charge ordering depend on the doping concentration.
Our calculations suggest that the observed
checkerboard patterns may well be induced by the long range Coulomb repulsion.

We consider a generalized $t-J$ model with an additional
long-range electron Coulomb potential,
\begin{eqnarray*}
H&=& H_t+H_s+H_c\nonumber\\
H_t& =& -t\sum_{\langle i,j\rangle\sigma}c^{\dag}_{i\sigma}c_{j\sigma}+hc \nonumber\\
H_s& =& J\sum_{\langle i,j\rangle}\mathbf{S}_{i}\cdot \mathbf{S}_{j}\nonumber\\
H_c& =& \frac{1}{2\epsilon}\sum_{i \neq j}\frac{\hat{n}_i
\hat{n}_j}{r_{ij}}, \label{hamiltonian}
\end{eqnarray*}
where $c_{i\sigma}$ is an annihilation operator of a spin $\sigma$
electron at site $i$. The sums in the kinetic and spin-exchange
Hamiltonian $H_t$ and $H_s$ run over all the nearest neighbor
pairs. $n_i=\sum_{\sigma}n_{i\sigma}$, and
$n_{i\sigma}=c^{\dag}_{i\sigma}c_{i\sigma}$. The sum in $H_c$ runs
over all the sites of $i$ and $j$. $\epsilon$ is the dielectric
constant, and $r_{ij}$ is the spatial distance between the two
sites $i$ and $j$. An positive charge background to balance the
charge neutrality is implied. There is a local constraint on every
site, $\sum_{\sigma}c^{\dag}_{i\sigma}c_{i\sigma} \leq 1$. In this Hamiltonian,
$H_c$ favors a charge ordering, while $H_t$ prefers a uniform charge
distribution.

We use the RMFT to estimate the energies of the charge ordered RVB
states. The RMFT was developed for the $t-J$ model to study a
charge homogeneous RVB state~\cite{zhang88,vanilla}. In that
theory, one considers a Gutzwiller projected BCS state for the
possible superconducting (SC) ground state. Here we shall extend
it to the charge inhomogeneous case. We consider a Gutzwiller
projected state,
\begin{eqnarray}
\braket{\Psi} = P_G \braket{\Psi_0},
\end{eqnarray}
where $\braket{\Psi_0}$ is a charge ordered BCS state, and $P_G =
\Pi_i (1-n_{i\uparrow}n_{i\downarrow})$ is the Gutzwiller
projection operator. We use Gutzwiller's approximation to relate
the expectation values of the kinetic or spin exchange energies in
the projected state $\braket{\Psi}$ (denoted by $< >$) to the
corresponding expectation values in the unprojected state
$\braket{\Psi_0}$ (denoted by $< >_0$) by two different
renormalization factors $g_t$ and $g_s$:
\begin{eqnarray}
\langle c^{\dag}_{i\sigma}c_{j\sigma}\rangle
   &\approx &  g_t^{ij}\langle c^{\dag}_{i\sigma}c_{j\sigma}\rangle_0 \nonumber\\
\langle \vec S_i \cdot \vec S_j\rangle
  & \approx &  g_s^{ij} \langle \vec S_i \cdot \vec S_j\rangle_0.
\end{eqnarray}
The renormalization factors are determined by the ratio
of the probabilities of the physical process in the
projected  and in the unprojected
states~\cite{zhang88,vanilla,gutzwiller,vollhardt}.
Similar to the method used~\cite{zhang88} for the homogeneous case, we find
\begin{eqnarray}
g_{t}^{ij}&=& 2\sqrt{\frac{(1-n_i)(1-n_j)}{(2-n_i)(2-n_j)}}, \nonumber\\
g_{s}^{ij}&=& \frac{4}{(2-n_i)(2-n_j)}. \label{gtgs}
\end{eqnarray}
They depend on the electron densities at the sites $i$
and $j$. In the homogeneous case, $n_i =n$,  $g's$ are
independent of the sites, we recover the results in
Ref.~\cite{zhang88}, $g^{(0)}_{t}=2x/(1+x)$ and
$g^{(0)}_{s}=4/(1+x)^2$, with $x=1-n$ the hole
density. The variational calculation of the
projected state $\braket{\Psi}$ in $H$ is then
mapped onto the unprojected state $\braket{\Psi_0}$ in
a renormalized Hamiltonian $H_{eff}$, given by
\begin{eqnarray}
H_{eff}&=& H'_t + H'_s + H_c\nonumber\\
H'_t &=& -t\sum_{\langle i,j\rangle \sigma}g_t^{ij}c^{\dag}_{i\sigma}c_{j\sigma}+ h.c.\nonumber \\
H'_s &=& J\sum_{\langle i,j\rangle}g^{ij}_s \mathbf{S}_{i}\cdot
\mathbf{S}_{j}.
\end{eqnarray}
Note that the intersite Coulomb interaction is not renormalized in
the theory ({\it i.e.} the renormalization factor is $1$).

Similar to the procedure in Ref.~\cite{zhang88}, we
introduce two mean fields: a particle-hole amplitude field
$\xi_{ij}=\sum_{\sigma}\langle c^{\dag}_{i\sigma}c_{j\sigma}\rangle_0$,
and a particle-particle pairing field
$\Delta_{ij} = \langle c_{i\uparrow}c_{j\downarrow}-c_{i\downarrow}c_{j\uparrow}\rangle_0$.
The renormalized Hamiltonian can then be solved by a self-consistent
mean field theory. The energy of $H_{eff}$ in the
unprojected state, hence the energy of the
generalized $t-J$ model in the projected state can be
written in terms of the self-consistent mean fields,
\begin{eqnarray}
E = - \sum_{\langle i,j\rangle}\bigl[2tg_t^{ij}\xi_{ij}
 +\frac{3J}{8}g_s^{ij}(\xi^2_{ij}+\Delta^2_{ij})\bigr] +
  \sum_{i \neq j} \frac{n_i n_j}{2\epsilon r_{ij}}.
\end{eqnarray}
In the  uniformly charged dRVB state, $\xi_{ij}=\xi$, and
$\Delta_{ij} = \pm \Delta$.  The energy per site is found to be
$E^{(0)}= -4tg^{(0)}_t\xi -(3J/4)g^{(0)}_s(\xi^2 +\Delta^2)$,
where we have dropped out the long range Coulomb energy of a
uniform electron density for it cancels to the energy due to the
oppositely charged background.

\begin{figure}
\includegraphics[width=4.8cm]{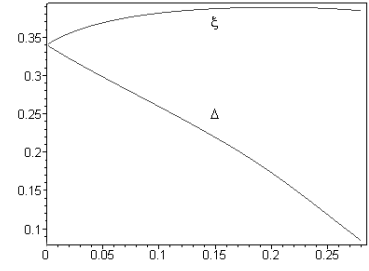}
\caption{\label{ki}The mean
fields $\Delta$ and $\xi$
versus the hole
concentration $x$ in the
uniform dRVB state.
$t/J=3$.}
\end{figure}

In the inhomogeneous case, the self-consistent equations, or the
Bogoliubov de-Genes equations, are more complicated. In what
follows we shall make an approximation to replace the mean fields
$\xi_{ij}$ and $\Delta_{ij}$ by their average mean values obtained
in the uniform dRVB state, and consider the effect of charge
ordering on the kinetic and spin-exchange energies due to the
renormalization factors $g_t^{ij}$ and $g_s^{ij}$, and on the
Coulomb potential. This is a rather dramatic approximation similar
to what proposed by Anderson~\cite{anderson}, but it should
capture a substantial part of the effect of the charge ordering.
The accuracy of this approximation will be examined
in a limiting case, which turns out to be quite good.
Within this approximation, the energy per site of the charge ordered dRVB
state relative to the uniform dRVB state is,
\begin{eqnarray}
\Delta E &=& \Delta E_t + \Delta E_s + \Delta E_c \nonumber\\
\Delta E_t &=& (\bar{g}_t - g^0_t) \langle H_t \rangle_0/N_s \nonumber\\
\Delta E_s &=& (\bar{g}_s - g^0_s \langle H_s \rangle_0/N_s \nonumber\\
\Delta E_c &=& e^2/(2\epsilon N_s)\sum_{ij}(n_i n_j-n^2)/r_{ij}.%
\label{energy-exression}
\end{eqnarray}
In the above equations, $\bar{g}_{t,s} = \sum_{\langle ij \rangle}
g^{ij}_{t,s}/2N_s$, $\langle H_{t,s} \rangle_0$ is the average
kinetic (spin exchange) energy in the uniform dRVB state. In
practice, we first solve the RMFT for the uniform dRVB state, from
which we obtain  $\xi$, $\Delta$, and $\langle H_{t,s}
\rangle_0$. In Fig.\ref{ki}, we plot the two mean fields as
functions of hole doping $x$ for $J/t=1/3$. We then calculate
$\bar{g}_{t,s}$ and $\Delta E_c$ for various types of charge
ordering patterns to estimate the energy of the charge ordered RVB
state, and to determine the optimal charge distribution. The
calculation of the long range Coulomb energy is similar to the
calculation of the Modelung constant, and it converges rapidly
with the appropriate choice of the method in summation.

Motivated by the approximate $4\times 4$
charge ordered states observed in STM experiments, we consider
several types of parent patterns shown below with
periodicity of $4a$ along both directions in the
square lattice. Each symbol represents a lattice site,
and the sites marked with the same symbol have the
same electron density.
\[
\setlength{\unitlength}{4mm}
\begin{picture}(18,5)(3,0)\linethickness{0.4pt}
\put(2,1) {\dashbox{0.05}(4.1,4.1)[l]
 {
   \put(0.3,3){$\circ$}\put(1.3,3){$\star$}\put(2.3,3){$\diamond$}\put(3.3,3){$\star$}
   \put(0.3,2){$\star$}\put(1.3,2){$\circ$}\put(2.3,2){$\star$}\put(3.3,2){$\diamond$}
   \put(0.3,1){$\diamond$}\put(1.3,1){$\star$}\put(2.3,1){$\circ$}\put(3.3,1){$\star$}
   \put(0.3,0){$\star$}\put(1.3,0){$\diamond$}\put(2.3,0){$\star$}\put(3.3,0){$\circ$}
 }
 \put(-2,-1){{\scriptsize I}}
}
\put(7,1)
{\dashbox{0.05}(4.1,4.1)[l]
 {
   \put(0.3,3){$\circ$}\put(1.3,3){$\star$}\put(2.3,3){$\diamond$}\put(3.3,3){$\star$}
   \put(0.3,2){$\star$}\put(1.3,2){$\circ$}\put(2.3,2){$\star$}\put(3.3,2){$\circ$}
   \put(0.3,1){$\diamond$}\put(1.3,1){$\star$}\put(2.3,1){$\circ$}\put(3.3,1){$\star$}
   \put(0.3,0){$\star$}\put(1.3,0){$\circ$}\put(2.3,0){$\star$}\put(3.3,0){$\circ$}
 }
 \put(-2,-1){{\scriptsize I\!I}}
}
\put(12,1)
{\dashbox{0.05}(4.1,4.1)[l]
 {
   \put(0.3,3){$\circ$}\put(1.3,3){$\star$}\put(2.3,3){$\star$}\put(3.3,3){$\star$}
   \put(0.3,2){$\circ$}\put(1.3,2){$\star$}\put(2.3,2){$\diamond$}\put(3.3,2){$\star$}
   \put(0.3,1){$\circ$}\put(1.3,1){$\star$}\put(2.3,1){$\star$}\put(3.3,1){$\star$}
   \put(0.3,0){$\circ$}\put(1.3,0){$\circ$}\put(2.3,0){$\circ$}\put(3.3,0){$\circ$}
 }
 \put(-2,-1){{\scriptsize I\!I\!I}}
}
\put(17,1)
{\dashbox{0.05}(4.1,4.1)[l]
 {
\put(0.3,3){$\circ$}\put(1.3,3){$\star$}\put(2.3,3){$\star$}\put(3.3,3){$\star$}
   \put(0.3,2){$\circ$}\put(1.3,2){$\star$}\put(2.3,2){$\diamond$}\put(3.3,2){$\star$}
   \put(0.3,1){$\circ$}\put(1.3,1){$\star$}\put(2.3,1){$\star$}\put(3.3,1){$\star$}
   \put(0.3,0){$\diamond$}\put(1.3,0){$\circ$}\put(2.3,0){$\circ$}\put(3.3,0){$\circ$}
 }
 \put(-2,-1){{\scriptsize I\!V}}
}
\label{p44}
\end{picture}
\]
We denote $n_\diamond=n +\delta_1$,
$n_\circ=n+\delta_2$, and
$n_\star=n+\delta_3$. Since
the overall average electron density of the
system is $n$, only two out of the three $\delta's$ are
independent. By using Eq.~(4), we have
$\bar{g}^I_{t,s}=\frac{1}{2}g_{t,s}^{\diamond\star}+\frac{1}{2}g_{t,s}^{\circ\star}$,
$\bar{g}^{I\!I}_{t,s}=\frac{1}{4}g^{\diamond\star}_{t,s}+\frac{3}{4}g^{\circ\star}_{t,s} $,
$\bar{g}^{I\!I\!I}_{t,s}=\frac{1}{4}g^{\star\star}_{t,s}+\frac{1}{4}g^{\circ\circ}_{t,s}
  +\frac{3}{8}g^{\star\circ}_{t,s}+\frac{1}{8}g^{\star\diamond}_{t,s}$,
  and
$\bar{g}^{I\!V}_{t,s}=\frac{1}{4}g^{\star\star}_{t,s}+\frac{1}{8}{g^{\circ\circ}_{t,s}}
+\frac{1}{8}g^{\star\diamond}_{t,s}+\frac{1}{8}{g^{\circ\diamond}_{t,s}}+\frac{3}{8}%
g^{\star\circ}_{t,s}$, here
the superscript in
$\bar{g}$ indicates the
type of the parent pattern,
and the superscript in $g$
refers to the two sites
with the marked symbols.
The Coulomb energy can be
shown to be quadratic in
$\delta's$, and they are
given by, in unit of
$e^2/\epsilon a$,
\begin{eqnarray}
&&\Delta E^{I}_c =-4.039\delta_1^2-4.039\delta_2^2-4.847\delta_1\delta_2\nonumber\\
&&\Delta E^{I\!I}_c =-1.497\delta_1^2-7.959\delta_2^2-3.468\delta_1\delta_2\nonumber\\
&&\Delta E^{I\!I\!I}_c =-0.567\delta_1^2-3.772\delta_2^2-2.511\delta_1\delta_2\nonumber\\
&&\Delta E^{I\!V}_c
=-1.248\delta_1^2-4.138\delta_2^2-1.428\delta_1\delta_2.
\label{coulomb energy}
\end{eqnarray}
Using these expressions, we have optimized the energy by
varying parameters $\delta_1$ and $\delta_2$,
and obtained charge ordered states with lower
energies. These states are derivatives of the parent
patterns under the consideration, but may have
a higher symmetry than the parent state for those
sites marked with different symbols may have the same
electron density. Below we shall discuss our results
in three different regions of the hole concentration.
In all of our calculations, we use $J/t=1/3$, $t=0.3
eV$, $a=3.8 \AA$.

\begin{figure}
\parbox{2.3cm}
{
\includegraphics[width=2cm]{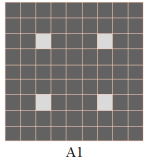}
}
\parbox{5.0cm}
{
 \caption{\label{fig:a1}\footnotesize
Low energy charge ordering pattern $A1$ of $4a \times 4a$ symmetry at hole density $x$
around 1/16. Shown are $9\times 9$ patches. Each square represents a lattice site and
the light (dark) square represnts low (high) electron density.}
}
\end{figure}

At the hole density around 1/16, the lower energy charge ordering
pattern is $A1$  as shown in Fig.\ref{fig:a1}. There are only two
types of the distinct sites in terms of the electron density in
this pattern. The numerical values of the energy gain and the
charge distributions are given in Table \ref{tab:1/16} for
$x=1/16$ and $x=0.05$. All other patterns at these dopings have
energies either higher than or too close to the energy of the
uniform dRVB state ($\Delta E >-0.01 eV$), and are not listed
here. At $x=1/16$, and $\epsilon=1$, the lowest energy state has a
charge distribution slightly deviated from a commensurate state
with the light site completely empty ($n=0$) and the dark site
fully occupied ($n=1$).  At a larger $\epsilon$, the energy gain
decreases rapidly and the energy of the pattern $A1$ is just
slightly lower than that of the homogenous case at $\epsilon=1.5$.
There is no stable charge ordering pattern at $\epsilon$ much
larger than $1.5$. $x=1/16$ is an ideal hole density for the pattern $A1$,
which was also discussed
in \cite{fu} and suggested in the magnetic and optical measurements~\cite{kim,zhou}.
At $x=0.05$, the pattern $A1$ is stable only
for smaller $\epsilon$, but is no longer stable for $\epsilon
=1.5$. Note that  the pattern $A1$  at
$x=0.05$ is an insulator for there
is no any connected path for holes to move through the
lattice.
\begin{table}\caption{\label{tab:1/16}\footnotesize
The energy and charge distribution of pattern $A1$ at hole density
$x=0.0625$ and $x=0.05$. An ideal hole crystal state with
$n_{\textcolor[rgb]{0.38,0.38,0.38}{\scriptscriptstyle{\blacksquare}}}=1$
at $x=0.0625$ is also listed for comparison.}
{\renewcommand{\arraystretch}{.96}
 \footnotesize
 \begin{tabular}{|c|c|c|c|c|}\hline
 $x$        &\multicolumn{3}{c|}{ $0.0625$ }  &$0.05$
  \\
   \cline{1-5}
 $\epsilon$ &\multicolumn{2}{c|}{ $1$ }    &  $1.5$  & $1$ \\
   \cline{1-5}
 $\Delta E$($eV$)
   &-0.038  &-0.037  & -0.010               & -0.016 \\
 $n_{\textcolor[rgb]{0.85,0.85,0.85}{\scriptscriptstyle{\blacksquare}}}$
   & 0.005  &0.000  & 0.017                & 0.200 \\
 $n_{\textcolor[rgb]{0.38,0.38,0.38}{\scriptscriptstyle{\blacksquare}}}$
   & 0.999  &1.000  & 0.999                & 1.000  \\
  Pattern
   & A1\,   &A1\,   & A1\,                 & A1\,   \\
  \hline
\end{tabular}
  }
\end{table}

\begin{table*}\caption{
\label{table0.125} The energy and charge distribution of lower
energy charge ordering patterns at hole density $x=0.125$ and $x=0.1$. The
energies of some patterns with charge distribution of
$n_{\textcolor[rgb]{0.38,0.38,0.38}{\scriptscriptstyle{\blacksquare}}}=1$
are also listed for comparison.} %
\begin{ruledtabular}
{\renewcommand{\arraystretch}{.96}
\footnotesize
\begin{tabular}{|c|c|c|c|c|c|c|c|c|c|c|c|c|c|c|c|c|c|}
$x$ &\multicolumn{10}{c|}{$0.125$}&\multicolumn{7}{c|}{$0.1$}\\
  \hline
$\epsilon$ &\multicolumn{5}{c|}{$1$} & \multicolumn{3}{c|}{$1.5$}
& \multicolumn{2}{c|}{$2$}
      &\multicolumn{5}{c|}{$1$}&\multicolumn{2}{c|}{$1.5$} \\
  \hline
$\Delta E $($eV$) & -0.116 &-0.113  &\,-0.056\, &-0.041\,
  &-0.013\,&-0.048 &-0.043&-0.025 &-0.015 & -0.010 &-0.059
  &-0.056&-0.051&-0.039&-0.021      &-0.021&-0.017 \\
   \hline
$n_{\scriptscriptstyle\square}$
  & \,   &\,   & 0.000 & 0.000 & \,  & \,   &\,  & 0.000  & \,   & 0.000  & \,   &
  \, & 0.000 & 0.000          & 0.000     & 0.000 &  \,   \\
$n_{\textcolor[rgb]{0.85,0.85,0.85}{\scriptscriptstyle\blacksquare}}$
  &0.006 &0.000 & 0.875 & 0.857 &0.542 & 0.018 &0.000 & 0.898  & 0.042   & 0.909  & 0.208 &
  0.200
  & 0.933 & 0.914 & 0.925         & 0.943 & 0.230  \\
$n_{\textcolor[rgb]{0.38,0.38,0.38}{\scriptscriptstyle{\blacksquare}}}$
  &0.999 &1.000 & 0.984 & 1.000 & 0.986  & 0.999 &1.000& 0.964  & 0.994& 0.954      & 0.999 &
  1.000
  & 0.984 & 1.000 & 1.000         & 0.975 & 0.996 \\
pattern
  &B1    & B1   &  B2  &  B2 &  B3  &  B1  &B1    &  B2   &  B1  &  B2   &  B1  &
  B1
  & B2   & B2   &  B4         &  B2  &  B1
\end{tabular}
}
\end{ruledtabular}
\end{table*}
\begin{table*}
\caption{
\label{tab0.15} Energies and charge distributions of
lower energy charge ordering patterns at $x=0.15$ }%
\begin{ruledtabular}
{\renewcommand{\arraystretch}{.96}
\footnotesize
\begin{tabular}{|c|c|c|c|c|c|c|c|c|c|c|c|c|c|c|c|c|c|c|}
  $\epsilon$& \multicolumn{8}{c|}{$1$}&\multicolumn{4}{c|}{$1.5$} &
  \multicolumn{4}{c|}{$2$}&$2.5$
   \\
   \hline
 $ \Delta E$($eV$)
       & -0.129  & -0.113 & -0.113 & -0.110 & -0.093 & -0.062 & -0.044 &
       -0.035
       & -0.063  & -0.048 & -0.046 & -0.044 &-0.029 &-0.148 &-0.011 &-0.010 &-0.010  \\
        \hline
 $n_{\scriptscriptstyle\square}$
       & \,  & 0.000   & 0.000   & 0.000   & 0.000   & 0.000   &  0.000   & \,   &
       \,
       & 0.000   & 0.000   & 0.000   & \,   & 0.000   & 0.000   &  0.000   & \, \\
 $n_{\scriptscriptstyle\textcolor[rgb]{0.85,0.85,0.85}{\blacksquare}}$
       & 0.000      & 0.933   & 0.950   & 0.950   & 0.933   & 0.828   &  0.800   & 0.428
       &0.000
       & 0.950   & 0.933   & 0.950   & 0.000      & 0.950   &  0.933   & 0.950   & 0.000      \\
 $n_{\textcolor[rgb]{0.38,0.38,0.38}{\scriptscriptstyle{\blacksquare}}}$
       & 0.972    & 1.000   & 1.000   & 1.000   & 1.000   & 0.975   & 1.000  &
       0.991
       & 0.973    & 1.000   & 1.000   &1.000&0.973 &1.000 &1.000 &1.000 & 0.973  \\
 Pattern
       &  B1    & C1     &  C3    & C2     & C4    & B2     & B2     &
       B3
       &  B1     & C3     &  C1    & C2     & B1   & C3     & C1     & C2   & B1\\
\end{tabular}
}
\end{ruledtabular}
\end{table*}
\begin{figure}
\includegraphics[width=2cm]{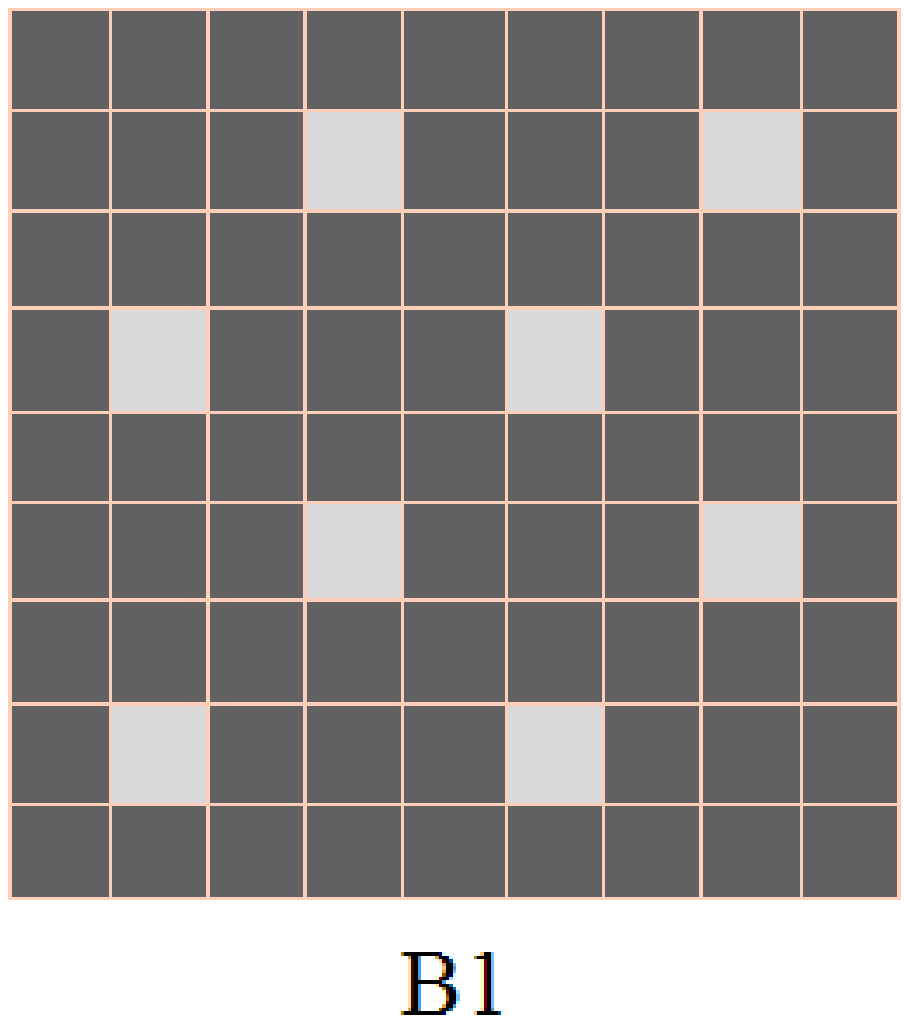}
\includegraphics[width=2cm]{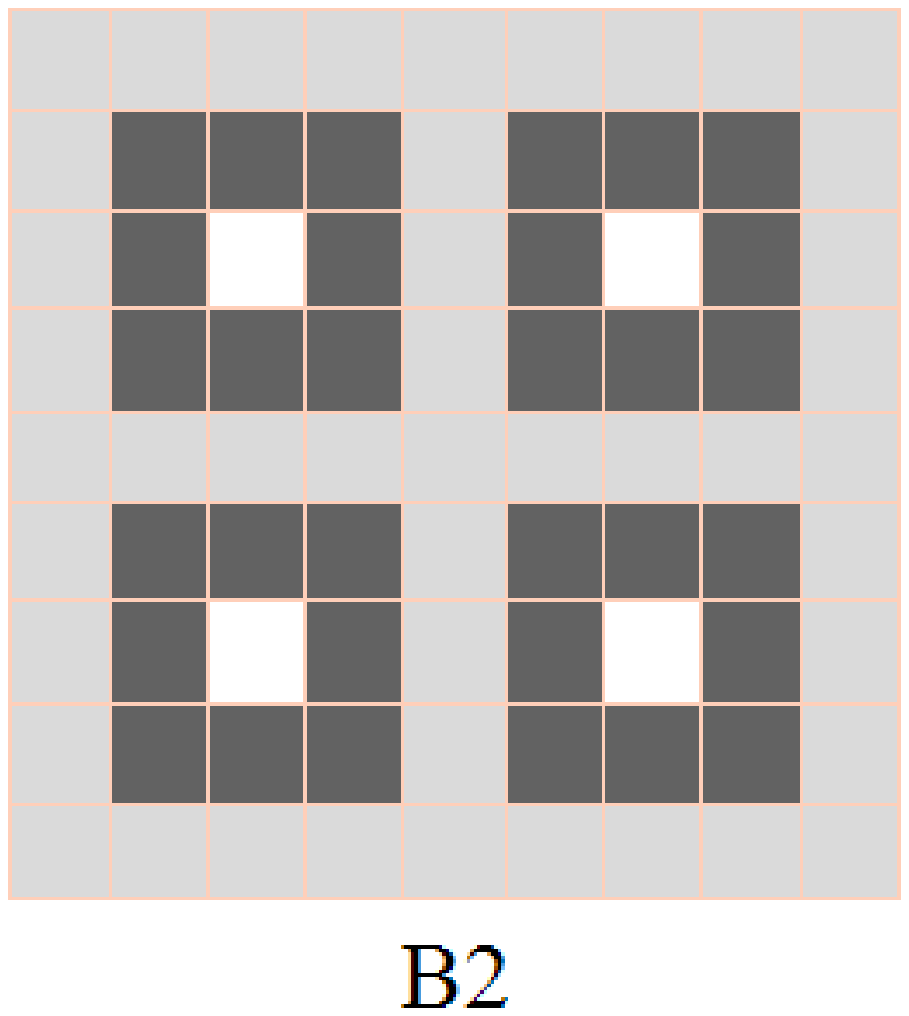}
\includegraphics[width=2cm]{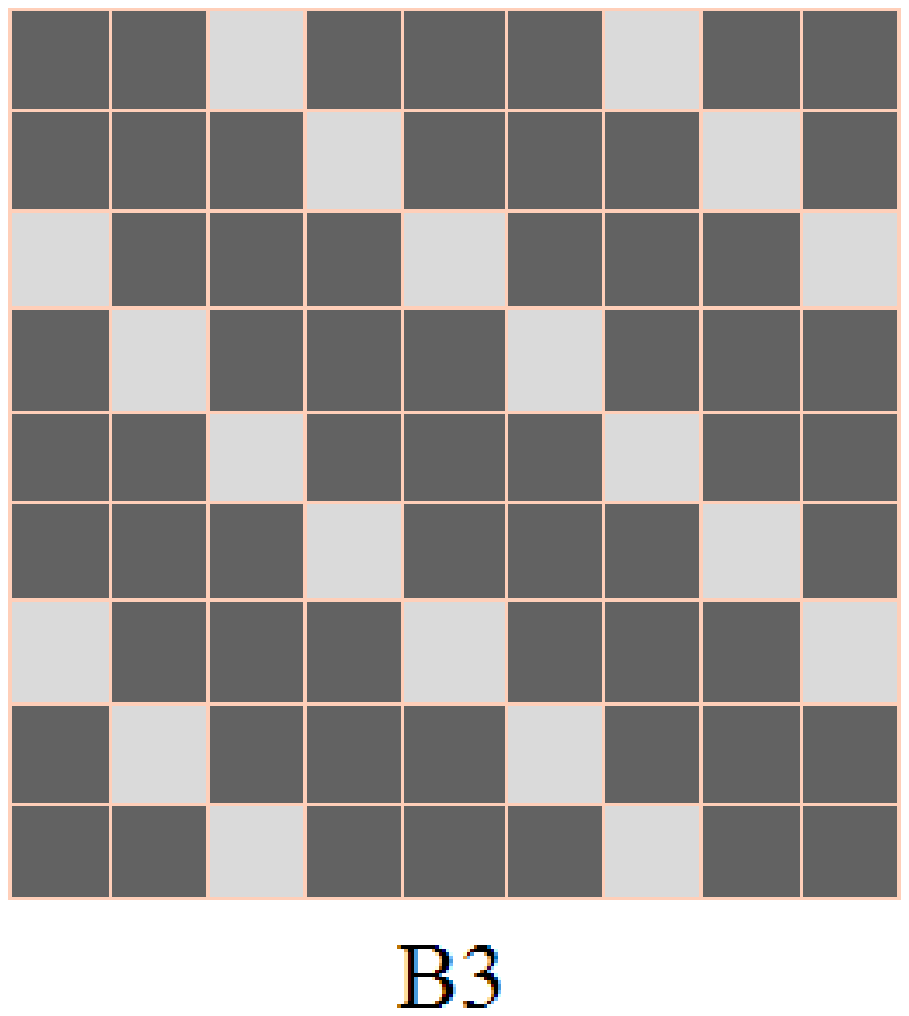}
\includegraphics[width=2cm]{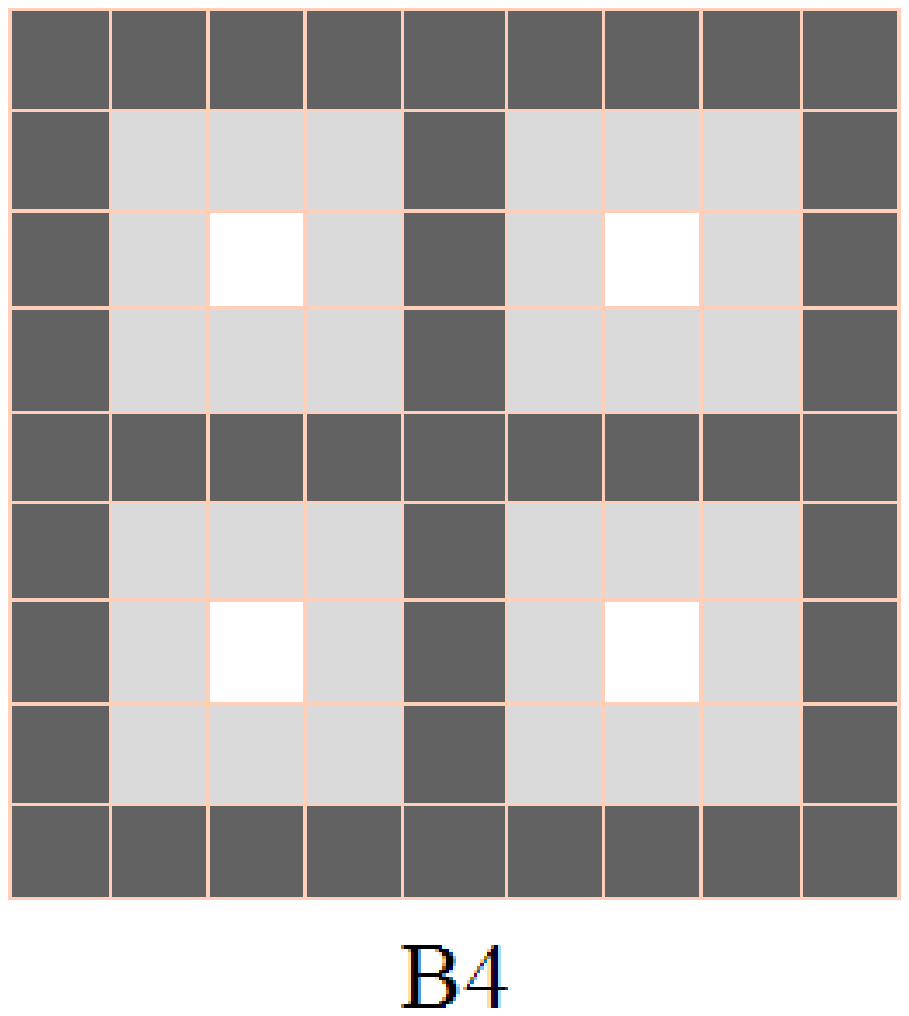}
\caption{\footnotesize
Lower energy charge ordering patterns for hole density around $x=1/8$.
$B1$ is of a symmetry of $\sqrt 8 a \times \sqrt 8 a$, and $B3$ is a stripe.}
\label{newpattern}
\end{figure}

At the hole density around $1/8$, there are several
charge ordering patterns as shown in Fig.\ref{newpattern}. Among
them the favorable  pattern is $B1$. Patterns $B2$ and
$B4$ have three types of distinct sites in terms of
the electron density, while $B1$ and $B3$ have two
types of distinct sites. The energy gain and the
charge distribution are given in Table \ref{table0.125} for
$x=1/8$ and $x=0.1$. Here we only list those patterns
with relatively lower energies. As we can see from table
\ref{table0.125}, at $x=1/8$ the energies of patterns $B1$ and $B2$ are
slightly lower than that of the homogeneous case at
$\epsilon=2$.  At $x=0.1$, the energy gain due to the
charge ordering at $\epsilon=2$ is already very tiny.

It is interesting to note that around the low hole density $x=1/8$,
both the checkerboard pattern $B2$ and the stripe pattern\cite{tranquada}
$B3$ are SC states for holes in these patterns can move through the lattice.

\begin{figure}
\includegraphics[width=2cm]{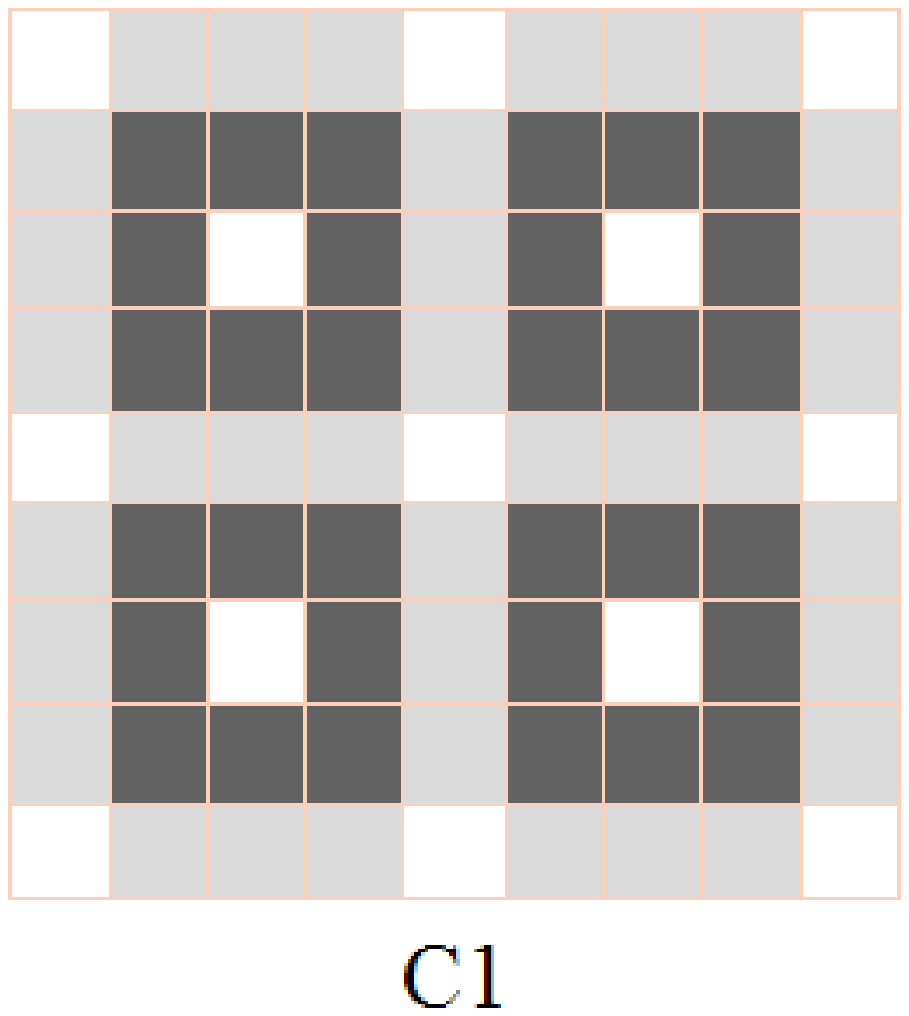}
\includegraphics[width=2cm]{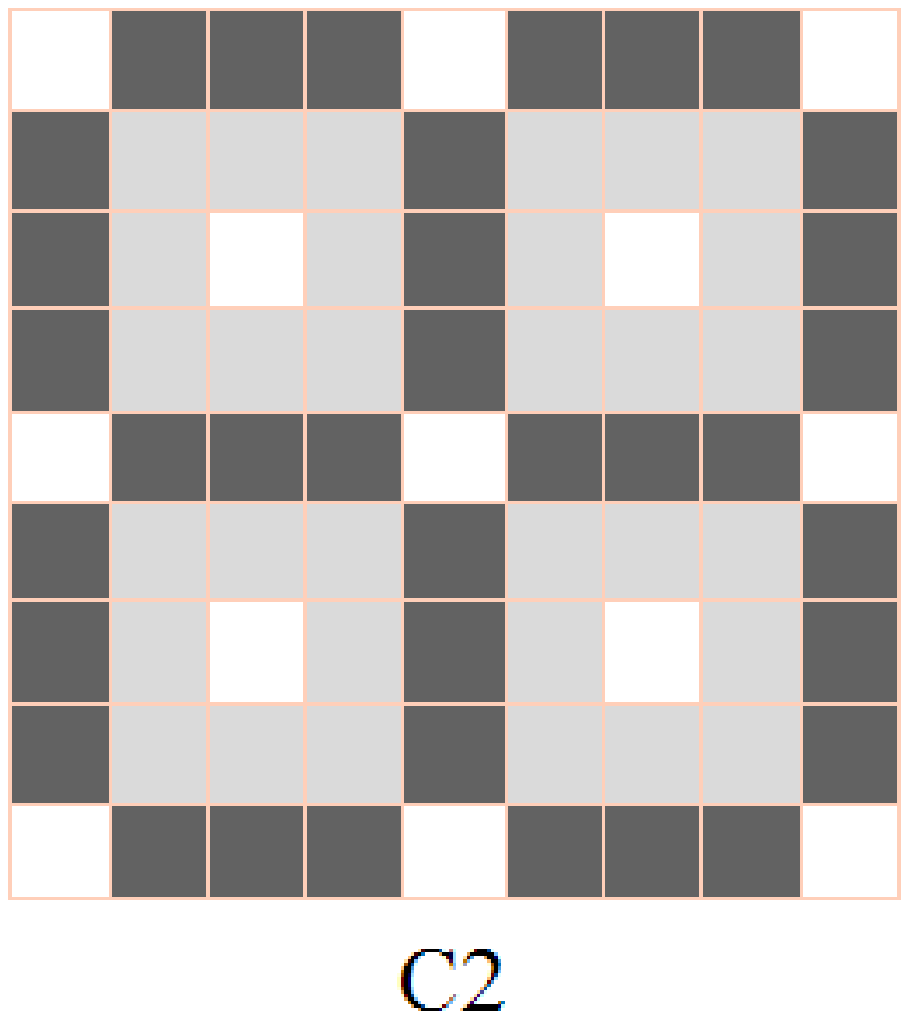}
\includegraphics[width=2cm]{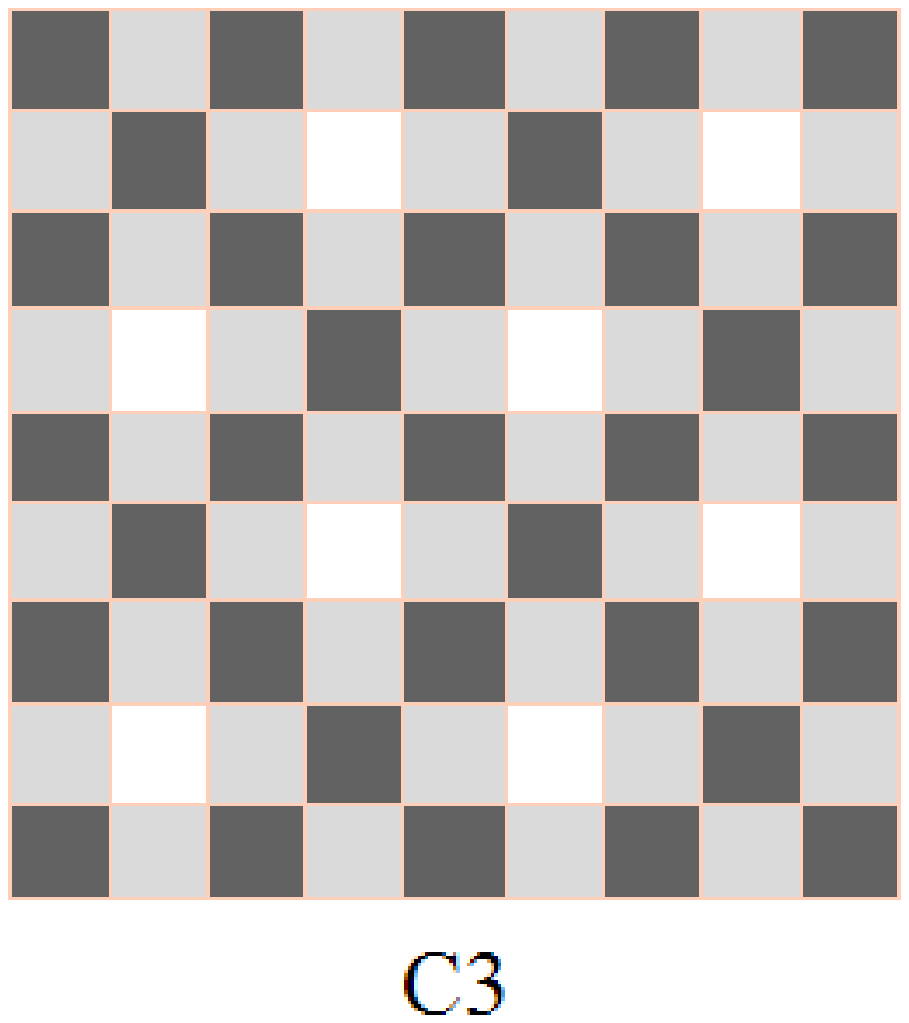}
\includegraphics[width=2cm]{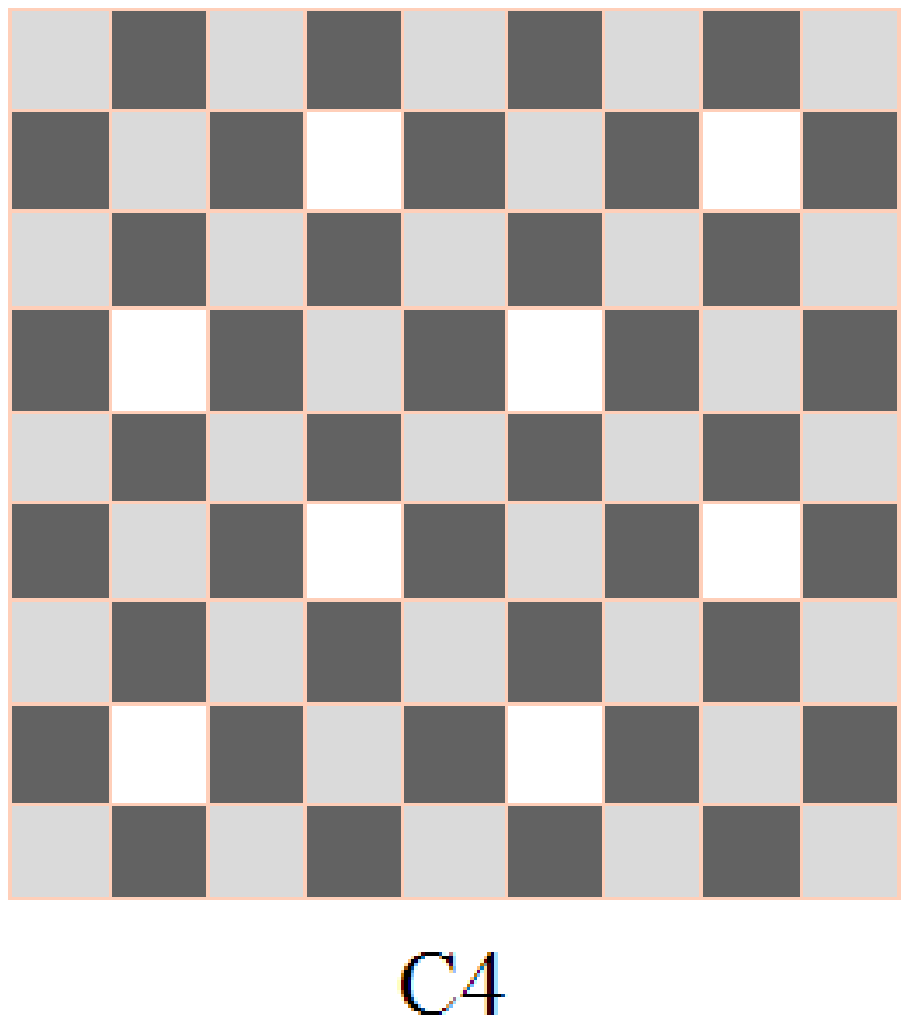}
\caption{\footnotesize
Several lower energy charge ordering patterns at
$x=0.15$, not included in Figs 3. Patterns C1 and C2
are related by the interchange of the dark and gray sites,
so for patterns C3 and C4.}
\label{lastp}\end{figure}

At high hole concentrations, several new charge ordering patterns
with lower energies appear, which are shown in Fig.\ref{lastp}. In Table
\ref{tab0.15}, we list the energies and charge densities of the lower
energy patterns at $x=0.15$. For $\epsilon=1$, the five patterns
($B1$, $C1$, $C2$, $C3$, and $C4$) have very close energies. In the pattern
C's, the electron densities at the dark and grey sites
are quite close. The empty sites in patterns $B1$ and C's form a
$\sqrt{8a}\times\sqrt{8a}$ Wigner hole crystal. Among the series C, pattern C1
is a conducting phase. We do not find any lower energy charge ordering
pattern at $\epsilon>2.5$. At $x=0.2$, the most favroable patterns are
$C1$, $C4$ and the stripe pattern $B3$.

In the energy estimation for the charge ordered RVB
state, we have focused on the effect of the charge
density dependent renormalization factors,
but neglected the site-dependence of the mean
fields $\xi$ and $\Delta$. This approximation turns
out to be quite good in a limiting case where all the
holes are completely localized at a single site.
Consider pattern $A1$ at $x=1/16$ and pattern $B1$
at $x=1/8$ with the electron density $n$ either
zero or 1. In this limit, the kinetic energy
vanishes. The spin exchange energy of the state can be
estimated by a direct counting of the missing
bonds due to the vacancies in an otherwise half filled
background, which is given by $E_s = 2(1-2x)\alpha $
per site, with $\alpha=-0.344 J$ the spin exchange energy per bond at
the half filling. For $J=0.1 eV$ we have $E_s =-0.060 eV$ at $x=1/16$ and
$ E_s = -0.052 eV$ at $x=1/8$, which are very close to the results
obtained in the present MFT: $E_s = 0.063 eV$ at $x=1/16$ and $E_s =-0.054
eV$ at $x=1/8$.

In summary we have studied the charge ordered RVB states in the doped
cuprates within a generalized $t-J$ model by using a renormalized mean
field theory. While the kinetic energy favors a uniform charge
distribution, the long range Coulomb repulsion tends to spatially modulate
the charge density in favor of charge ordered RVB states. Since both the
Coulomb potential and the leading order in kinetic
energy are quadratic in the density variation, we
expect and indeed have found that the charge
density variation from the uniform state is always
large in the charge ordered state. The stability of
the charge ordered RVB state strongly depends on
the dielectric constant $\epsilon$. In cuprates,
$\epsilon \approx 2.5 - 5$. Our calculation suggests
that the observed charge ordered state in STM
experiments in cuprates may well be interpreted due to
the long range Coulomb interaction. Among the
favorable charge ordered superconducting states,
pattern $B1$ has a symmetry of $\sqrt 8 \times \sqrt
8$, patterns $B2$ and $C1$ both have checkerboard
structure, and pattern $B3$ is a stripe. Because of the
intersite Coulomb repulsion, we do not find
the bound hole paris in the charged ordered states.

The work is partially supported by NSFC No.10225419 and 90103022,
and by RGC in Hong Kong, and by the US NSF ITR grant No. 0113574.
One of us (FCZ) thanks P. W. Anderson for providing Ref.~\cite{anderson}
prior to the publication.


\begin{references}
\bibitem{hoffman1} J.E. Hoffman, {\it et al.}, Science {\bf 295}, 466 (2002).
\bibitem{howald} C. Howald {\it et al.}, Phys. Rev. B {\bf 67}, 014533 (2003).
\bibitem{hoffman2}J.E. Hoffman, {\it et al.}, Science {\bf 297}, 1148 (2002).
\bibitem{mcelroy1} K. McElroy {\it et al.}, Nature {\bf 422}, 520 (2003).
\bibitem{vershinin}
M. Vershinin {\it et al}, Science {\bf 303}, 1995 (2004)
\bibitem{mcelroy2} K. McElroy {\it et al.}, to be published (2004).
\bibitem{hanaguri} T. Hanaguri {\it et al.}, submitted to Nature.
\bibitem{franz} M. Franz, D. E. Sheehy, and Z. Tesanovic, Phys. Rev. Lett. {\bf 88}, 257005 (2002).
\bibitem{wang} Q. Wang and D. H. Lee, Phys. Rev. {\bf B67}, 020511 (2003).
\bibitem{chen} H.D. Chen, O. Vafek, A. Yazdani, and S.C. Zhang,
cond-mat/0402323.%
\bibitem{fu} H.C. Fu, J.C. Davis and D-H Lee, cond-mat/0403001.
\bibitem{anderson} P.W. Anderson, cond-mat/0406038.%
\bibitem{laughlin} R. Laughlin, LANL con-mat/0209269.
\bibitem{zhang03} F. C. Zhang, Phys. Rev. Lett. {\bf 90}, 207002 (2003).
\bibitem{zhang88} F.C. Zhang, C. Gros, T.M. Rice and H. Shiba,
J. Supercond. Sci. Tech.{\bf 1}, 36 (1988). %
\bibitem{vanilla} P. W. Anderson, P. A. Lee, M. Randerai, N. Trievedi, and F. C. Zhang,
J. Phys.: Condensed Matter {\bf 24} R755, (2004).
\bibitem{gutzwiller} M. C. Gutzwiller, Phys. Rev. {\bf 137}, A1726 (1965).
\bibitem{vollhardt} D. Vollhardt, Rev. Mod. Phys. {\bf 56}, 99 (1984).
\bibitem{kim} Y. H. Kim and P. H. Hor, Mod. Phy. Lett. {\bf B15}, 497 (2001);
P. H. Hor and Y. H. Kim, J. Phys.: Condens. Matter {\bf 14}, 10377 (2002).
\bibitem{zhou} F. Zhou {\it et al.}, Supercon. Sci. Technol. {\bf 16}, L7 (2003)
\bibitem{tranquada} J.M. Tranquada {\it et al},
Nature {\bf 375}, 561 (1995); J. Zaanen, O. Gunnarson, Phys. Rev.
B. {\bf 40}, 7391 (1989);
V.J. Emery, S.A. Kivelson, Physica C {\bf 209}, 597 (1993).
\end{references}
\end{document}